%% file: ppg10.tex
\documentclass[11pt]{elsart}
\usepackage{color}
\usepackage{epsfig}
\usepackage{rotating,lscape}
\usepackage{url}
\usepackage{cite}

\newcommand\fverb{\setbox\pippobox=\hbox\bgroup\verb}
\newcommand\fverbdo{\egroup\medskip\noindent%
            \fbox{\unhbox\pippobox}\ }
\newcommand\fverbit{\egroup\item[\fbox{\unhbox\pippobox}]}
\newbox\pippobox
\newcommand{\affuni}[2]{Dipartimento di Fisica dell'Universit\`a #1, #2, Italy.}
\newcommand{\affinfnm}[2]{INFN Sezione di #2, #2, Italy.}
\newcommand{\affinfnn}[2]{INFN Sezione di #1, #2, Italy.}

\def\ifm#1{\relax\ifmmode#1\else$#1$\fi}
  
    \def\dif{\hbox{d\kern.1mm}}
    
\def\pic{\ifm{\pi^+\pi^-}} \def\epm{\ifm{e^+e^-}}  \def\to{\ifm{\rightarrow\,}}
\def\plm{\ifm{\pm}}    \def\deg{\ifm{^\circ}}  
\def\gam{\ifm{\gamma}}  \def\x{\ifm{\times}}  \def\ab{\ifm{\sim}}  
\def\ie{{\it\kern-1pt i.\kern-.5pt e.\kern-.2pt}}  
\let\cl=\centerline    
\def\lfigbox#1;#2;{\parbox{#2cm}{\vglue3mm\epsfig{file=#1.eps,width=#2cm}\vglue3mm}}
\def\km{\kern-1.5mm}  \def\kak{\km&\km}  \def\kma{\kern-2.5mm}\def\kms{\kern-.75mm}
\def\pt#1,#2,{\ifm{#1\x10^{#2}}}   \def\bye{\end{document}}
\def\strike#1;{\setbox0=\hbox{#1}\rlap{\vrule height3.3pt depth-2.3pt width\wd0}\kern2pt#1\ }

\def\lfigbox#1;#2;{\parbox{#2cm}{\vglue3mm\epsfig{file=#1.eps,width=#2cm}\vglue3mm}}

\catcode`@=11 
\newdimen\z@ \z@=0pt 
\newskip\z@skip \z@skip=0pt plus0pt minus0pt
\def\m@th{\mathsurround=\z@}
\def\ialign{\everycr{}\tabskip\z@skip\halign} 
\def\eqalign#1{\null\,\vcenter{\openup\jot\m@th
  \ialign{\strut\hfil$\displaystyle{##}$&$\displaystyle{{}##}$\hfil
      \crcr#1\crcr}}\,}
\catcode`@=12 
\newcommand{\eV}{{e\kern-.07em V}}

\newcommand{\dd}{\ensuremath{{\rm d}}}

\definecolor{dmagen}{rgb}{.7,.0,.7}
\definecolor{dgree}{rgb}{.0,.5,.0}
\definecolor{dred}{rgb}{.9,.0,.0}

\makeatletter
\def\@ptsize{1}
\belowdisplayskip \abovedisplayskip
\abovedisplayshortskip \z@ \@plus 2\p@%
\def\section{\@startsection{section}{1}{.5\z@}{1\@bls
  \@plus .1\@bls \@minus .1\@bls}{.5\@bls}{\normalsize\bfseries}}
\def\subsection{\@startsection{subsection}{2}{.5\z@}{.3\@bls
  \@plus .1\@bls \@minus .1\@bls}{.3\@bls}{\normalsize\itshape}}
\makeatother

\begin{document}

\begin{frontmatter}
  \title{\mathversion{bold} Measurement of 
    $\sigma(e^+e^-\rightarrow\pi^+\pi^-)$
    from threshold to 0.85 GeV$^2$ 
    using Initial State Radiation 
    with the KLOE detector}\vspace{-0.5cm}
  \collab{The KLOE Collaboration}
  \author[Na,infnNa]{F.~Ambrosino},
\author[Roma2,infnRoma2]{F.~Archilli},
\author[Mainz]{P.~Beltrame\corauthref{cor}\thanksref{cern}}\ead{beltrame@kph.uni-mainz.de},
\author[Frascati]{G.~Bencivenni},
\author[Roma1,infnRoma1]{C.~Bini},
\author[Frascati]{C.~Bloise},
\author[Roma3,infnRoma3]{S.~Bocchetta},
\author[Frascati]{F.~Bossi},
\author[infnRoma3]{P.~Branchini},
\author[Frascati]{G.~Capon},
\author[Frascati]{T.~Capussela},
\author[Roma3,infnRoma3]{F.~Ceradini},
\author[Frascati]{P.~Ciambrone},
\author[Frascati]{E.~De~Lucia},
\author[Roma1,infnRoma1]{A.~De~Santis},
\author[Frascati]{P.~De~Simone},
\author[Roma1,infnRoma1]{G.~De~Zorzi},
\author[Mainz]{A.~Denig\corauthref{cor}}\ead{denig@kph.uni-mainz.de},
\author[Roma1,infnRoma1]{A.~Di~Domenico},
\author[infnNa]{C.~Di~Donato},
\author[Roma3,infnRoma3]{B.~Di~Micco},
\author[Frascati]{M.~Dreucci},
\author[Frascati]{G.~Felici},
\author[Roma1,infnRoma1]{S.~Fiore},
\author[Roma1,infnRoma1]{P.~Franzini},
\author[Frascati]{C.~Gatti},
\author[Roma1,infnRoma1]{P.~Gauzzi},
\author[Frascati]{S.~Giovannella},
\author[infnRoma3]{E.~Graziani},
\author[Frascati]{M.~Jacewicz},
\author[Karlsruhe]{W.~Kluge},
\author[Frascati,StonyBrook]{J.~Lee-Franzini},
\author[Karlsruhe]{D.~Leone},
\author[Na,infnNa]{P.~Massarotti},
\author[Na,infnNa]{S.~Meola},
\author[Frascati]{S.~Miscetti},
\author[Mainz]{S.~M\"uller\corauthref{cor}\thanksref{kvi}}\ead{muellers@kph.uni-mainz.de},
\author[Frascati]{F.~Murtas},
\author[Na,infnNa]{M.~Napolitano},
\author[Roma3,infnRoma3]{F.~Nguyen},
\author[infnRoma3]{A.~Passeri},
\author[Frascati,Energ]{V.~Patera},
\author[Frascati]{P.~Santangelo},
\author[Roma3,infnRoma3]{C.~Taccini},
\author[infnRoma3]{L.~Tortora},
\author[Frascati]{G.~Venanzoni},
\author[Frascati,Energ]{R.Versaci}
\address[Frascati]{Laboratori Nazionali di Frascati dell'INFN, 
Frascati, Italy.}
\address[Karlsruhe]{Institut f\"ur Experimentelle Kernphysik, 
Universit\"at Karlsruhe, Germany.}
\address[Mainz]{Institut f\"ur Kernphysik, 
Johannes Gutenberg - Universit\"at Mainz, Germany.}
\address[Na]{Dipartimento di Scienze Fisiche dell'Universit\`a 
``Federico II'', Napoli, Italy.}
\address[infnNa]{INFN Sezione di Napoli, Napoli, Italy.}
\address[Energ]{Dipartimento di Energetica dell'Universit\`a 
``La Sapienza'', Roma, Italy.}
\address[Roma1]{\affuni{``La Sapienza''}{Roma}}
\address[infnRoma1]{\affinfnm{``La Sapienza''}{Roma}}
\address[Roma2]{\affuni{``Tor Vergata''}{Roma}}
\address[infnRoma2]{\affinfnn{Roma Tor Vergata}{Roma}}
\address[Roma3]{\affuni{``Roma Tre''}{Roma}}
\address[infnRoma3]{\affinfnn{Roma Tre}{Roma}}
\address[StonyBrook]{Physics Department, State University of New York
  at Stony Brook, USA.}
\vspace{2.2cm}
\begin{abstract}
We have measured the cross section of the radiative process $e^+e^- \to \pi^+\pi^-\gamma$ with the KLOE detector at the Frascati $\phi$-factory DA$\Phi$NE, from events taken at a CM energy $W$=$1$ GeV. Initial state radiation allows us to obtain the cross section for $e^+e^- \to \pi^+\pi^-$, 
the pion form factor $|F_\pi|^2$ and the dipion contribution to the muon magnetic moment anomaly, $\Delta a_\mu^{\pi\pi} = 
(478.5\pm2.0_\mathrm{stat}\pm5.0_\mathrm{syst}\pm4.5_\mathrm{th})\times
10^{-10}$ in the range $0.1 < M_{\pi\pi}^2 < 0.85$ GeV$^2$, where the 
theoretical error includes a \mbox{SU(3) $\chi$PT} estimate of the uncertainty on 
photon radiation from the final pions.
The discrepancy between the
Standard Model evaluation of $a_\mu$ and the value
measured by the Muon g-2 collaboration at BNL is confirmed.  
\end{abstract}\vglue-5mm
\vglue-5mm
\corauth[cor]{\noindent Corresponding Authors}
\thanks[cern]{Now at UCLA Physics and Astronomy Dept., Los Angeles, California
90095-1547, USA.}
\thanks[kvi]{Now at KVI, 9747 AA Groningen, The Netherlands.}
\end{frontmatter}
\overfullrule6pt

\section{Introduction}
\label{sec:Intro}
The anomaly of the magnetic moment of the muon, $a_\mu=(g_\mu-2)/ 2$, is one of the best measured quantities in particle physics~\cite{Bennett:2006fi}. Recent 
evaluations of the hadronic contributions to the anomaly~\cite{Jegerlehner:2009ry,Davier:2009zi} 
lead to a discrepancy of about 3 standard deviations of the Standard Model (SM) value from the result of the 
Brookhaven $(g_\mu-2)$ experiment~\cite{Bennett:2006fi}. A large part of the 
uncertainty of the
theoretical estimate comes from the leading order hadronic
contribution $\Delta a_\mu^{\mathrm{had,lo}}$, which at low energies is not calculable by perturbative QCD, but can be
evaluated via a dispersion relation using measured 
cross sections of $e^+e^- \to$ hadrons~\cite{disp_int}. Initial state 
radiation (ISR) allows to obtain these cross sections at
$e^+e^-$ colliders operating at fixed energies~\cite{actis:2009gg}, from the production threshold up to the collision energy. 
The energy region below 1 GeV, which 
is accessible with the KLOE experiment at DA$\Phi$NE in Frascati, is dominated by the $\pi^+\pi^-$ 
channel and contributes
$\sim 75\%$ to the value of $\Delta a_\mu^{\mathrm{had,lo}}$, and accounts for $\sim 40\%$ 
of its total uncertainty~\cite{Jegerlehner:2009ry}. Better accuracy for the dipion cross
section results in an improvement of the SM prediction for $a_\mu$.

The KLOE collaboration has already published two measurements of the dipion cross
section for $M_{\pi\pi}^2$ between 0.35 and 0.95 GeV$^2$ using $e^+e^-\to \pi^+\pi^-\gamma$ events from data collected in 2001~\cite{plb606} and 2002~\cite{plb670}, both at a collision energy $W$=$M_\phi$.
We present in the following a new measurement, based on data taken in 2006 at $W = 1$ GeV, about 20 MeV below the $\phi$-meson mass, using
different acceptance criteria for the radiated photons. In our previous
measurements, the photon was required to be
emitted at small polar angles ($\theta < 15^0$ or $\theta > 165^0$) with respect to the beamline, and therefore
escaped detection. In the measurement presented in this Letter, we
require the photon to be detected in the
electromagnetic calorimeter of KLOE at large polar angles. This allows to
extend the $M_{\pi\pi}^2$ region down to the threshold for the dipion production.

\section{Measurement of the cross section $e^+e^- \to \pi^+\pi^-$ at DA$\Phi$NE}
\label{sec:Meas}
The KLOE experiment operates at the Frascati $\phi$-factory
DA$\Phi$NE, an $e^+e^-$-collider with beams crossing at $\pi - 0.025$ radians running 
mainly at a center-of-mass energy $W \simeq 1020$ MeV, the $\phi$-meson mass. 
The 
DA$\Phi$NE collision energy can be changed only marginally away
from the $\phi$-resonance energy, and measurements 
of hadronic cross sections scanning a wider energy range are not
possible. 
However, events with photons radiated by the initial state electron or positron producing a $\pi^+\pi^-$ pair can cover energies from threshold up to the collision energy. 
KLOE measures the differential cross section  for $e^+ e^-\to\pi^+\pi^-\gamma$ as a function of the $\pi^+\pi^-$
invariant mass squared, $M^2_{\pi\pi}$, and from this obtains the dipion cross section
$\sigma_{\pi\pi}\equiv\sigma(e^+ e^-\to\pi^+\pi^-)$ according to~\cite{Binner:1999bt}:
\begin{equation}
\frac{\dd\sigma(e^+e^-\to\pi^+\pi^-\gamma)} {\dd M_{\pi\pi}^2} =
\frac{\sigma_{\pi\pi}(M_{\pi\pi}^2)}{s}~ H(M_{\pi\pi}^2,s)~.
\label{eq:1}
\end{equation}
Eq. \ref{eq:1} defines the dimensionless ``radiator function''
$H$, which can be obtained from QED calculations. Since there is no way to distinguish ISR photons from final state radiation (FSR) photons in the KLOE detector, corrections are necessary and are properly included in the analysis. 

\begin{figure}[t!]
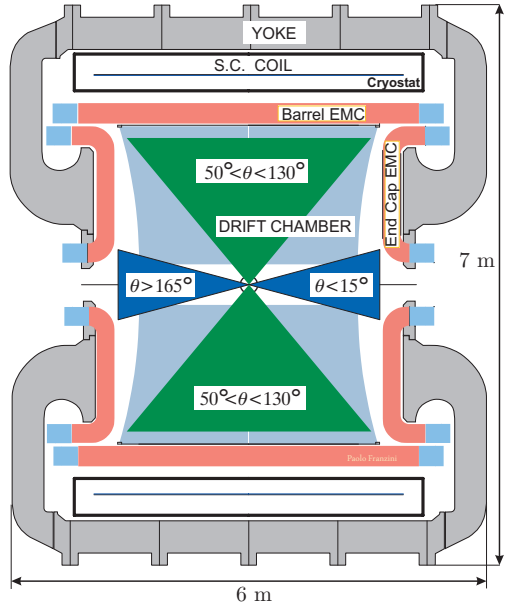

\cl{\lfigbox pfkloe;6.6;}
\caption{Vertical cross section in the $y-z$ plane of the KLOE detector, showing the small and large angle regions for photons and pions used in the different KLOE measurements.}
\label{fig:det}
\end{figure}

The KLOE detector, see
Fig.~\ref{fig:det}, consists of a cylindrical drift chamber~\cite{Adinolfi:2002uk}
surrounded by an electromagnetic calorimeter (EMC)~\cite{Adinolfi:2002zx}. A superconducting coil provides a magnetic field of 0.52 T along the $z$-axis.\footnote{The line which bisects the angle between the two colliding beams is
  taken as the $z$-axis of the KLOE coordinate system with incoming positrons
  going along positive values of $z$. The $x$-axis is
horizontal, pointing to the center of the collider rings, while the
$y$-axis is vertical, directed upwards.}
The drift chamber measures track points with a
resolution of \ab~0.15 mm in $r$-$\phi$ and \ab~2 mm in $z$. The momentum resolution is
$\sigma_{p_t}/p_t\sim 0.4\%$ for tracks with polar angle
$45^\circ<\theta<135^\circ$. 
Energy deposits in the calorimeter close in space and time are
combined in ``clusters'' by the reconstruction program. 
The cluster energy resolution is $\sigma_E/E\sim 5.7\%/\sqrt{E\ {\rm(GeV)}}$ and the time
resolution is $\sigma_t\sim 54\ {\rm ps}/\sqrt{E\ {\rm(GeV)}}\oplus
100\ {\rm ps}$. 

As mentioned above, the previous KLOE measurements~\cite{plb606,plb670} used events with photons emitted
within cones of $\theta_\gamma<15^\circ$ around the 
beamline (narrow cones in Fig.~\ref{fig:det}) and two charged pion tracks 
 with $50^\circ<\theta_\pi<130^\circ$ (wide cones in Fig.~\ref{fig:det}). 
In this configuration,  the photon is
not detected, its direction
is reconstructed from the pion momenta
by closing kinematics: $\vec{p}_\gamma\simeq\vec{p}_\mathrm{miss}= -(\vec{p}_{\pi^+}
+\vec{p}_{\pi^-})$. These requirements guarantee high
  statistics for ISR events (because of the divergence of the ISR cross section at small photon angles), and a reduced background contamination (from the resonant process $e^+e^-\to
  \phi\to\pi^+\pi^-\pi^0$ 
as well as from the final state 
radiation process $e^+e^-\to \pi^+\pi^-\gamma_\mathrm{FSR}$).
However, requiring the photons at small angles, the low mass dipion region is not reachable since below $0.35$ GeV$^2$ both pions are emitted at small angles and therefore outside acceptance, resulting in a loss of events.   

To reach the dipion threshold, in the new measurement 
we require events to have a photon detected in the calorimeter 
at large polar angles, 
$50^\circ<\theta_\gamma<130^\circ$ (wide cones in
Fig.~\ref{fig:det}), the same region where also pion tracks are detected. 
However, compared to 
the measurements with photons at small angles, these conditions imply a
reduction in statistics of about a factor of 5, and an increase of the background from the process $\phi \to \pi^+\pi^-\pi^0$, as well as the
irreducible background from events with final state radiation and from $\phi$ radiative 
decays. The 
hadronic 
uncertainties associated with the theoretical description  
 of the $\phi$ radiative decays to the scalar mesons $f_0(980)$ and 
$f_0(600)$ together with the background from the $\phi\to\rho\pi\to(\pi\gamma)\pi$ decay 
contribute to the uncertainty of the measurement~\cite{Ambrosino:2005wk}. 
To reduce the background contamination and the mentioned uncertainties, we collected data at a collision energy of $W=1$ GeV, 
4.5 $\Gamma_\phi$ (about 20 MeV) below the $\phi$-meson peak,
decreasing the $\phi$-meson production by about a factor of 80. 
This reduces the effect of $f_0\gamma$ and $\rho\pi$ decays of the
$\phi$-meson to the level of a few percent.

\subsection{Event selection}
\label{sec:Meas_1}
\begin{figure}[t!]
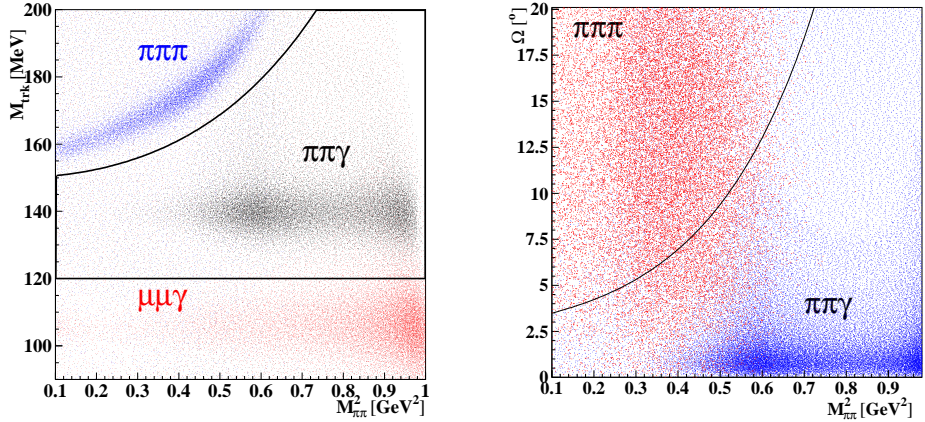

\cl{\lfigbox trkmcuts_mc;5.6;\kern1cm\lfigbox omecuts_mc;5.6;}
\caption{Left: MC signal and background distributions in the $M_{\rm
    trk}$--$M^2_{\pi\pi}$ plane. Right: the same, in the $\Omega$--$M^2_{\pi\pi}$ plane. Black lines indicate the cuts described in the text.}
\label{fig:kincuts}
\end{figure}

Requirements to select events are:
\begin{enumerate}
 \item A trigger from two energy deposits larger than 50
  MeV in two sectors of the barrel calorimeter~\cite{Adinolfi:2002hs}.
  \item A fast offline background filter has to be satisfied~\cite{Ambrosino:2004qx} to reject machine background and cosmic ray events.  
\item Two tracks with opposite sign curvature and satisfying $50^\circ<\theta <130^\circ$ coming from the interaction point. The latter condition is obtained requiring each track 
   to cross a cylinder
   centered around the interaction point with $8$ cm radius and $14$ cm
   length. Cuts on
   $|\vec{p}|>200$ MeV and $p_t>160$ MeV or $|p_z|>90$ MeV, respectively, are
   required to ensure good reconstruction efficiency
  \item At least one photon with
    $50^\circ<\theta_\gamma<130^\circ$
    and $E_\gamma > 20$ MeV must be detected, where a photon is defined as a cluster in the EMC not associated to a track. If several photons fulfill the criteria, the one with the highest energy is chosen.
  \item A particle identification variable ($\pi$-e PID) is evaluated for each track associated to a cluster 
    in the calorimeter, and an
    event with both tracks identified as electrons, due to radiative Bhabha scattering events, is
    rejected. 
  \item The event must satisfy cuts on the track mass variable, $M_{\rm trk}$.\footnote{$M_{\rm trk}$ is computed from the measured
      momenta of the two particles $\vec{p}_\pm$ assuming they have the same mass:
      $
      \left(\sqrt{s}-\sqrt{|\vec{p}_+|^2 + M^2_{\rm trk}}-
      \sqrt{|\vec{p}_-|^2 + M^2_{\rm trk} }\right)^2-\left(\vec{p}_+
      +\vec{p}_-\right)^2 = M_\gamma^2 = 0~.
      $
    } Fig.~\ref{fig:kincuts}, left, shows how a cut in $M_{\rm trk}>120$ MeV rejects
    $\mu^+\mu^-\gamma$ events, while a $M_{\pi\pi}^2$-dependent cut
    rejects $\pi^+\pi^-\pi^0$ events.   
  \item $\pi^+\pi^-\pi^0$ events are further rejected by a cut on the
    angle $\Omega$ between the directions of the detected photon and of the missing momentum
    $\vec{p}_\mathrm{miss}$. Fig.~\ref{fig:kincuts}, right, shows the
    $M_{\pi\pi}^2$-dependent cut used to reject $\pi^+\pi^-\pi^0$
    events situated at large $\Omega$ values. 
\end{enumerate}
About $0.6$ million events in the  $M_{\pi\pi}^2$ range between 0.1
and 0.85 GeV$^2$ are selected.   

\subsection{Determination of the cross section}
\label{sec:Meas_2}
The radiative differential cross section is obtained subtracting
the background count $N_{\rm bkg}$ from the observed count $N_{\rm obs}$ in bins of $\Delta M_{\pi\pi}^2=0.01$ GeV$^2$, and dividing by
the selection efficiency, $\epsilon(M_{\pi\pi}^2)$,
and by the integrated luminosity $\mathcal{L}$:
\begin{equation}
\frac{\dd\sigma_{\pi\pi\gamma}}
{\dd M_{\pi\pi}^2} = \frac{N_{\rm obs}-N_{\rm bkg}}
{\Delta M_{\pi\pi}^2}\, \frac{1}{\epsilon(M_{\pi\pi}^2)~ \mathcal{L}}\,~.
\label{eq:2}
\end{equation}

\subsubsection{Background subtraction}

After selection cuts, residual background events from
$\mu^+\mu^-\gamma$, $\pi^+\pi^-\pi^0$, $e^+e^-\gamma$ and a small fraction of 
$\phi\to
K^+K^-$, $\phi \to \eta\gamma$ events survive. Their number, $N_{\rm bkg}$, is found by fitting the
$M_{\rm trk}$ spectrum of the selected data sample with a superposition
of Monte Carlo (MC) distributions describing signal and
background (the $e^+e^-\gamma$ distribution is obtained from a
control sample of data using the $\pi$-e PID estimator to select
electrons). 
The fit parameters 
are the normalization factors for the background distributions,
obtained for 15 intervals in $M^2_{\pi\pi}$ of 0.05 GeV$^2$ width. The 
background contamination is dominated by the $\mu^+\mu^-\gamma$
contribution and is found to 
be less than 10\% above 0.3 GeV$^2$, while reaching the level of 50\%
at the dipion production threshold. Systematic uncertainties of the 
background estimates are obtained from the errors on the 
normalization coefficients, yielding values smaller than 0.2\% above 
0.5 GeV$^2$, with a gradual increase to 3.4\% at 
threshold. 

We also estimate the contribution from the processes
$e^+e^-\to \phi \to (f_0,\sigma)\gamma\to\pi^+\pi^-\gamma$ and 
$e^+e^-\to \rho^\pm\pi^\mp\to (\pi^\pm\gamma)\pi^\mp$ to the signal using a modified version of the PHOKHARA MC 
generator~\cite{phok61}. Despite the 
fact that the data have been taken with DA$\Phi$NE running 20 MeV below
the mass of the $\phi$ meson, an effect of several percent is found, mostly below 0.3 GeV$^2$, that 
needs to be subtracted from the spectrum. 
Moreover, there is also a sizable non-resonant $\rho^\pm\pi^\mp$ 
contribution.
The systematic error assigned to this 
contribution reflects the uncertainty of the production mechanism for these 
channels. It is negligible above 0.5 GeV$^2$, but reaches a value of 6.5\% 
at threshold. 
\subsubsection{Efficiency and systematics evaluation}
Efficiencies for the offline background filter, trigger and the $\pi$-e PID estimator
are obtained from data control samples. All other
efficiencies (including geometrical acceptance) are evaluated as one combined {\it
  global efficiency} from samples of MC generated events passing the
full simulation of the detector response. 

Events are generated using the PHOKHARA event generator, which includes
next-to-leading order ISR~\cite{Rodrigo:2001kf} and leading order FSR
calculations, as well as simultaneous emission of one ISR and one FSR
photon~\cite{Czyz:2003ue}. The generator is interfaced with the KLOE
detector simulation code GEANFI~\cite{Ambrosino:2004qx}. All MC 
efficiencies are compared to the efficiencies obtained from data
control samples, and small corrections are applied to the efficiencies for tracking and photon detection. For all other efficiencies, the MC prediction agrees well with the results from data.   
 
{\it Offline background filter.}\kern3mm Its efficiency is
evaluated from a downscaled control sample retained during the data
taking, and is larger than 99\%. To overcome statistical limitations of the control sample, a
polynomial parametrization is used below \mbox{0.4 GeV$^2$}. The uncertainty
of the parameters introduces a systematic error ranging from 0.1 to 0.5\%.

{\it Trigger.}\kern3mm The efficiency is obtained from
a subsample of $\pi^+\pi^-\gamma$ events in which two out of the three particles
satisfy the trigger requirements. The trigger response
for the third particle is parametrized as a function of its momentum and
direction, and the efficiency as a function of $M^2_{\pi\pi}$ is obtained
using kinematic event distributions from MC. It is larger than 99.5\%. As a consistency check, 
the procedure is applied to a sample of $\pi^+\pi^-\gamma$
events from MC and the outcome is compared to the MC efficiency for an event to
satisfy the trigger criteria using the same sample. The fractional difference between the two methods of a few per mill
is taken as the systematic uncertainty.

{\it Pion-electron PID.}\kern3mm The PID estimator is based on time-of-flight and energy and shape of the calorimeter cluster associated to each track. Each track is extrapolated  to the
calorimeter and at least one cluster is searched for within a sphere of radius
$|\vec{r}_{\rm imp}\!-\!\vec{r}_{\rm clu}|\!<\!$ 90 cm around the
track impact point, $\vec{r}_{\rm imp}$. 
The efficiency for each track is evaluated on a clean sample of
$\pi^+\pi^-\gamma$ events from data where one track
with an associated cluster is identified to be a pion, 
and evaluating the probability
for the other track to have an associated cluster and also to be recognized as a
pion. From this, the event probability to satisfy the selection
criteria of having at least one track
to be identified as a pion is found. It has been verified to be larger than 99.5\% using control samples from data and MC. A similar consistency check as in the trigger
efficiency evaluation reveals a maximum uncertainty of a few per mill only below
0.15 GeV$^2$.


{\it Tracking.}\kern3mm This efficiency is contained in the
global MC efficiency. Its value is between 97 and 98\%. 
The correction for the difference in data and MC
efficiency is obtained comparing the efficiencies for a single pion
track as a function of momentum and polar angle
from MC and data control samples containing a fully reconstructed
pion track of opposite charge and one photon. Event kinematics from MC
are then used to get the efficiencies as a function of $M^2_{\pi\pi}$.   
The data efficiency is found to be approximately $0.3\%$ lower than
the MC efficiency due to the presence of split tracks not well reproduced
in the simulation. The MC-data difference is included as a
correction in the analysis. The corresponding systematic uncertainty is estimated
varying the radius and length of the cylindrical region around the
interaction point the tracks must cross to be selected. It is found to be 
$0.3$\%.

{\it Photon detection.}\kern3mm The photon detection efficiency 
has been measured using a sample of $\pi^+\pi^-\pi^0$
events selected from data requiring two oppositely charged tracks and
one photon coming from the decay of a $\pi^0$. The efficiency is estimated requiring to observe 
a second photon in a cone
around the predicted direction. The efficiency results to be close to 100\%, 
and data and MC efficiencies are in excellent agreement in the energy range of interest.
The value from data is few per mill lower
only for $M_{\pi\pi}^2 > 0.8$ GeV$^2$. Therefore, the
systematic uncertainty is considered
negligible. 

The efficiencies for the cuts in $M_{\rm trk}$ and $\Omega$ as well as
the geometrical acceptance for pions and photon are already included in the
global efficiency from MC. Their systematic uncertainties are obtained
as follows:
\begin{itemize}
\item The systematic uncertainties due to the $M_{\rm trk}$ and $\Omega$
  cuts are obtained by varying the cuts shown in Fig.~\ref{fig:kincuts}
  within reasonable limits of the resolution in $M_{\rm trk}$ and
  $\Omega$ angle ($\sigma_{M_{\rm trk}}\sim 3$ MeV, $\sigma_\Omega\sim
  2^\circ$) and evaluating the effect on the $\pi^+\pi^-\gamma$
  spectrum. For $M_{\rm trk}$, one obtains an uncertainty that is in the range of 0.1 to 0.4\% above 0.5 GeV$^2$,  whereas below it increases to 3\% at  
threshold. The uncertainty on the $\Omega$ cuts is negligible above 0.5 GeV$^2$, and reaches 1.4\% at threshold.
 \item In a similar way, the systematic effects due to the polar angle
   requirements for the pions and the detected photon ($50^\circ <
   \theta_{\pi,\gamma}< 130^\circ$) are estimated by changing the angular
   acceptance by $\pm 2^\circ$ for $\theta_\pi$, and $\pm 5^\circ$ for
   $\theta_\gamma$. The resulting uncertainty is about 0.3\% above 0.5 GeV$^2$. 
   Below, it increases to 1.9\% at threshold.   
\end{itemize}
The detector resolution is unfolded using a Bayesian method~\cite{dagostini}. The high momentum resolution of the KLOE drift chamber makes this correction small, and as a result the statistical errors for different bins in $M_{\pi\pi}^2$ become only weakly correlated. Comparison of different unfolding methods gives non-negligible differences only in the two bins close to the $\rho-\omega$ interference ($0.60 < M_{\pi\pi}^2< 0.62$ GeV$^2$). The difference of about 2\% is taken as a systematic uncertainty for these two bins.

The absolute energy calibration of the KLOE detector is validated with a fit of the pion form factor~\cite{KLOEurl}. The $\omega$-meson mass is found to be $m_\omega = (782.6\pm0.3)$ MeV, in excellent agreement with the value from PDG~\cite{Amsler:2008zzb}.

Parametrized fractional systematic uncertainties as functions of $M_{\pi\pi}^2$ are given in~\cite{KLOEurl}. Fractional systematic uncertainties which are constant over the range of $M_{\pi\pi}^2$ covered in this measurement are listed in Table~\ref{tab:syseff2}. 

\subsubsection{Luminosity and radiative corrections}
\label{sec:Meas_3}
The absolute normalization of the data sample is performed by
counting Bhabha events at large polar angles ($55^\circ<\theta<125^\circ$).
The effective cross section is $\sigma_\mathrm{Bhabha}\simeq 430$ nb. To obtain the
integrated luminosity, $\mathcal{L}$, the observed number
of Bhabha events is divided by the
effective cross section evaluated by the Monte Carlo generator
Babayaga@NLO~\cite{CarloniCalame:2000pz,Balossini:2006wc},
which includes QED radiative corrections
with the parton shower algorithm, and which has been interfaced with
the KLOE detector simulation. The estimated theoretical uncertainty of this
generator is $0.1\%$. The experimental uncertainty on the 
luminosity measurement is 0.3\%, dominated by
the systematics on the angular acceptance. The integrated luminosity of the dataset used in the analysis is $(232.6\pm0.2_{\rm th}\pm0.7_{\rm exp})$ pb$^{-1}$, with negligible statistical error.
A detailed description of the KLOE luminosity measurement can be found
in~\cite{Ambrosino:2006te}.

The radiator function $H$ used to extract the cross
section $\sigma_{\pi\pi}$ from the measured differential cross section
for \epm\to\pic\gam~ in
Eq.~\ref{eq:1} is obtained from the PHOKHARA MC generator,
which includes complete next-to-leading order ISR
corrections~\cite{Czyz:2003}, with a precision of $0.5\%$ mostly due to the effect of missing 
higher order terms. 
In addition, the cross section is corrected
for the vacuum polarization~\cite{Jeger_alpha} (running of $\alpha_{\mathrm{em}}$), and the
shift between the measured value of $M^2_{\pi\pi}$ and the squared 
virtual photon 4-momentum transfer $q^2\equiv(M^0_{\pi\pi})^2$ 
for events with pions radiating a photon in the final 
state. Again the PHOKHARA generator, 
which includes FSR in the pointlike-pion approximation~\cite{Czyz:2005}, 
is used to estimate the second correction: a matrix relating 
$M^2_{\pi\pi}$ to 
$(M^0_{\pi\pi})^2$, giving the probability for an event in a bin of
$M^2_{\pi\pi}$ to originate from some different bin of
$(M^0_{\pi\pi})^2$, is used to correct the spectrum. 

The validity of the pointlike-pion approach used in the MC generator is compared
with a SU(3) $\chi$PT calculation~\cite{Fuchs:2000pn}. For intermediate and high 
values of $M_{\pi\pi}$ no significant disagreement is found, while below the $\rho$ mass peak region, deviations of up to 7\% at the two-pion threshold are found~\cite{Ivashyn:2009pi}.  In absence of more advanced theoretical 
investigations, we take the \mbox{$M_{\pi\pi}$-dependent} difference between 
the two methods 
as an estimate of the systematic uncertainty related to the pointlike-pion 
approach. The entry for ``FSR treatment'' in Table~\ref{tab:syseff2} takes 
into account this uncertainty, as well as the one due to the limited 
knowledge of the pion form factor value at 
$\sqrt{s}=1$ GeV.

\section{Results}
\label{sec:Results}
The differential $\pi^+\pi^-\gamma$ cross section is obtained from
Eq.~\ref{eq:2} performing the analysis as described in
Sec.~\ref{sec:Meas_2}. The total cross section $\sigma_{\pi\pi}$ is
then computed dividing by the radiator function $H$, as described in Eq.~\ref{eq:1}. To obtain the bare cross section, $\sigma_{\pi\pi}^{\rm bare}$, we remove the effects from 
vacuum polarization of the virtual photon produced in the $e^+e^-$ annihilation according to:
\begin{equation}
\sigma_{\pi\pi}^{\rm bare}(s') = \sigma_{\pi\pi}(s')\times \left(\frac{\alpha(0)}{\alpha(s')} \right)^2\;~, 
\label{eq:3}
\end{equation}   
where $s'\equiv (M_{\pi\pi}^0)^2$ and $\alpha(0)$ is the fine structure constant in the limit $q^2=0$ ($\alpha(0) = e^2 / 4\pi \epsilon_0 \hbar c$), and $\alpha(s')$ represents its effective value at $(M^0_{\pi\pi})^2$. We use the parameterization given in~\cite{vacpol} for ${\alpha(0)}/{\alpha(s')}$.   

The squared modulus of the pion form factor
$|F_\pi|^2$ is derived from
\begin{equation} 
|F_\pi(s')|^2 (1+\eta_{\rm FSR}(s'))=\frac{3}{\pi}\frac{s'}{\alpha^2
  \beta_\pi^3}\sigma_{\pi\pi}(s')\,~,  
\label{eq:4}
\end{equation}
where  $\beta_\pi = \sqrt{1-
{4m_\pi^2}/{s'}}$ is the pion velocity. The factor $(1+\eta_{\rm FSR}(s'))$ describes the effect of FSR assuming pointlike pions (see~\cite{Schwinger:1989ix,Jegerlehner:2006ju}). In this way, for the radiative corrections applied to $\sigma_{\pi\pi}^{\rm bare}$ and $|F_\pi|^2$, we adopt the same definition used in energy scan measurements~\cite{CMD2,CMD2_2,SND}: $\sigma_{\pi\pi}^{\rm bare}$ is inclusive with respect to final state radiation, and undressed from vacuum polarization effects, while $|F_\pi|^2$ contains vacuum polarization effects with final state radiation removed.

Our results are summarized in Table~\ref{tab:results}, which gives
\begin{itemize}
\item the observed differential cross section $\dd \sigma(e^+ e^-
  \rightarrow \pi^+ \pi^-\gamma)/\dd M_{\pi\pi}^2$ as a function of the measured invariant mass
 of the dipion system, $M^2_{\pi\pi}$, with $0^\circ < \theta_\pi < 180^\circ$ and at least one photon 
  in the angular region $50^\circ < \theta_\gamma < 130^\circ$ with $E_\gamma > 20$ MeV, with statistical and systematic error;
\item the {\it bare}  cross section  $\sigma^{\rm bare}(e^+ e^- \rightarrow \pi^+ \pi^-)$, inclusive of FSR, but 
with vacuum polarization effects removed, as a function of 
$(M^0_{\pi\pi})^2$, with statistical error;
\item the squared modulus of the pion form factor, dressed with vacuum polarization, but with FSR effects removed,
 as a function of $(M^0_{\pi\pi})^2$, with statistical error.
\end{itemize}
The statistical errors given in Table~\ref{tab:results} are weakly
correlated as a result of the resolution unfolding. 
The systematic error for each value of
$M^2_{\pi\pi}$ is obtained combining in quadrature  all the individual 
contributions in each column in Table~\ref{tab:syseff2}. For the differential 
cross section for \epm\to\pic\gam, these systematic errors are reported for 
convenience in Table~\ref{tab:results}. Polynomial parameterizations as a function of $M_{\pi\pi}^2$ can be found in~\cite{KLOEurl} if the 
contributions listed in Table~\ref{tab:syseff2} are not constant in $M^2_{\pi\pi}$. 

\begin{figure}[t!]
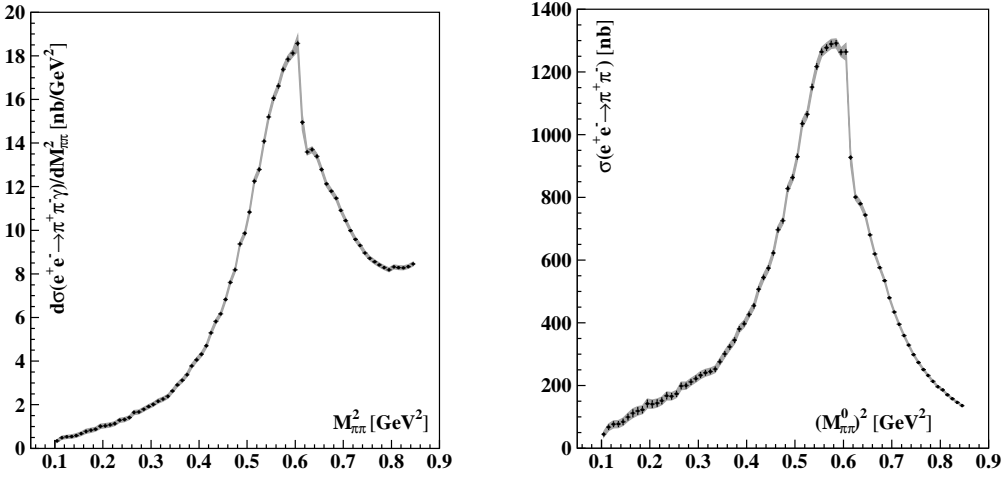
\centering
\lfigbox sppg09;6;\kern1cm\lfigbox spp09;6.3;
\caption{Left: Differential cross section for \epm\to\pic\gam,
with $50\deg < \theta_\gamma < 130\deg$. Right: bare cross section $\sigma_{\pi\pi}^{\rm bare}$ for $e^+e^-\to \pi^+\pi^-$. Data points have statistical error attached, the gray band gives the statistical and systematic uncertainty (added in quadrature).}
\label{fig:sigppg}
\end{figure}
Fig.~\ref{fig:sigppg}, left, shows the observed differential cross
section for \epm\to\pic\gam, while Fig.~\ref{fig:sigppg}, right,
shows the cross section  $\sigma^{\rm bare}_{\pi\pi}$. The latter is
the input to the dispersion integral for $\Delta a_{\mu}^{\pi\pi}$~\cite{disp_int}: 
\begin{equation}
\Delta a_{\mu}^{\pi\pi}=\frac{1}{4\pi^3}\int_{s_{min}}^{s_{max}}\dd
s\,\sigma^{\rm bare}_{\pi\pi}(s)\,K(s),
\label{amuint}
\end{equation}
which is computed as the sum of the values for $\sigma^{\rm bare}_{\pi\pi}$
listed in Table~\ref{tab:results} multiplied by the bin width of 0.01 GeV$^2$
 and the kernel function $K(s)$, which
  behaving approximately like $1/s$~\cite{deraf1} enhances the
  contributions at low
  values of $s$.
The integration limits are
$s_{min}=0.10$ GeV$^2$ and $s_{max}=0.85$ GeV$^2$.
Statistical errors of the $\sigma^{\rm bare}_{\pi\pi}$ values are summed in quadrature to obtain the 
statistical error of
$\Delta a_{\mu}^{\pi\pi}$. The systematic error of $\Delta
a_{\mu}^{\pi\pi}$ is obtained as follows:
the individual systematic uncertainties of the 
$\sigma^{\rm bare}_{\pi\pi}$ values (listed in Table~\ref{tab:syseff2}) are added linearly in the
summation because they are all fully bin-to-bin correlated. Then the different contributions to the systematic uncertainty of $\Delta
a_{\mu}^{\pi\pi}$ are added in
quadrature to get the total experimental and theory systematic errors.  
We find
\begin{equation}
\Delta a_\mu^{\pi\pi}(0.1-0.85\; {\rm GeV^2}) = (478.5 \pm
  2.0_{\rm stat} \pm 5.0_{\rm exp} \pm 4.5_{\rm th}) \times 10^{-10}~.
  \label{eqn:amupipi_pop}
\end{equation}
The combined fractional systematic error of our value for $\Delta
a_\mu^{\pi\pi}$ is 1.4\%. 

Data tables and covariance 
matrices as well as further documentation of the measurement are given in~\cite{KLOEurl}.

\begin{table}[htb!]
\begin{center}
\renewcommand{\arraystretch}{1.0}
\begin{tabular}{||l|c|c|c|c||}
\hline
 & $\sigma_{\pi\pi\gamma}$ & $\sigma_{\pi\pi}^{\rm bare}$ & $|F_\pi|^2$ & $\Delta a_\mu^{\pi\pi}$ \\
\cline{2-4}
 & \multicolumn{3}{|c|}{\begin{tabular}{c@{ \scriptsize{;} }c}\scriptsize{threshold}&\scriptsize{$\rho$-peak}\end{tabular}} & (0.1 - 0.85 GeV$^2$)\\
\hline
\hline
Background Filter & \multicolumn{3}{|c|}{\begin{tabular}{c@{ \scriptsize{;} }c} 0.5\%& 
0.1\%\end{tabular}} & negligible \\
\cline{5-5}
Background subtraction & \multicolumn{3}{|c|}{\begin{tabular}{c@{ \scriptsize{;} }c} 3.4\%&0.1\%\end{tabular}} & 0.5\% \\
\cline{5-5}
$f_0+\rho\pi$ bkg. & \multicolumn{3}{|c|}{\begin{tabular}{c@{ \scriptsize{;} }c} 6.5\%&
negl.\end{tabular}} & 0.4\% \\
\cline{5-5}
$\Omega$ cut & \multicolumn{3}{|c|}{\begin{tabular}{c@{ \scriptsize{;} }c} 1.4\%&
negl.\end{tabular}} & 0.2\% \\
\cline{5-5}
Trackmass cut & \multicolumn{3}{|c|}{\begin{tabular}{c@{ \scriptsize{;} }c} 3.0\%&
0.2\%\end{tabular}} &   0.5\% \\
\cline{5-5}
$\pi$-e PID & \multicolumn{3}{|c|}{\begin{tabular}{c@{ \scriptsize{;} }c} 0.3\%&
negl.\end{tabular}} & negligible\\
\cline{5-5}
Trigger & \multicolumn{3}{|c|}{\begin{tabular}{c@{ \scriptsize{;} }c} 0.3\%&
0.2\%\end{tabular}} & 0.2\%\\
\cline{5-5}
Acceptance & \multicolumn{3}{|c|}{\begin{tabular}{c@{ \scriptsize{;} }c} 1.9\%&
0.3\%\end{tabular}} & 0.5\%\\
\cline{5-5}
Unfolding & \multicolumn{3}{|c|}{\begin{tabular}{c@{ \scriptsize{;} }c} negl.& 
2.0\%\end{tabular}} & negligible \\
\cline{2-5}
Tracking & \multicolumn{4}{|c||}{0.3\%} \\
\cline{2-5}
Software Trigger (L3) & \multicolumn{4}{|c||}{0.1\%} \\
\cline{2-5}
Luminosity & \multicolumn{4}{|c||}{0.3\%} \\
\cline{2-5}\hline
Experimental syst. & \multicolumn{3}{|c|}{ } & 1.0\% \\\hline\hline
FSR treatment & - & \multicolumn{2}{|c|}{\begin{tabular}{c@{ \scriptsize{;} }c} 7\%& 
negl.\end{tabular}}  & {0.8\%} \\
\cline{2-5}
Radiator function $H$ & - & \multicolumn{3}{|c||}{0.5\%} \\
\cline{2-5}
Vacuum Polarization & - & Ref.~\cite{vacpol} & - & 0.1\% \\
\cline{2-5}\hline
Theory syst. & \multicolumn{3}{|c|}{ } & 0.9\% \\
\hline\hline
\end{tabular}
\caption{Systematic errors on $\sigma_{\pi\pi\gamma}$,
  $\sigma_{\pi\pi}^{\rm bare}$, $|F_\pi|^2$ and $\Delta a_\mu^{\pi\pi}$. 
All errors are fully bin-to-bin correlated. If the error is not constant over the range of $M_{\pi\pi}^2$, the value at threshold and at the $\rho$-peak (0.6 GeV$^2$) is given. The uncertainty on $\Delta a_\mu^{\pi\pi}$ is composed
of a $0.6\%$ contribution coming from the SU(3) $\chi$PT calculation and a $0.5\%$ one from the
uncertainty of $|F_\pi|^2$ at $\sqrt{s}=1$ GeV. Complete parameterizations of the errors can be found in Ref.~\cite{KLOEurl}.}
\label{tab:syseff2}
\end{center}
\end{table}

\begin{landscape}
\begin{table}[p]
\centering
\renewcommand{\arraystretch}{1.2}
{\scriptsize
\begin{tabular}{||c|c|c|c||c|c|c|c||c|c|c|c||}
\hline
\km$M^2_{\pi\pi}|(M^0_{\pi\pi})^2$\km&$\sigma_{\pi\pi\gamma}$&$\sigma_{\pi\pi}^{bare}$&\parbox{1cm}{$|F(\pi)|^2$\vglue-3mm}&%
\km$M^2_{\pi\pi}|(M^0_{\pi\pi})^2$\km&$\sigma_{\pi\pi\gamma}$&$\sigma_{\pi\pi}^{bare}$&\parbox{1cm}{$|F(\pi)|^2$\vglue-3mm}&%
\km$M^2_{\pi\pi}|(M^0_{\pi\pi})^2$\km&$\sigma_{\pi\pi\gamma}$&$\sigma_{\pi\pi}^{bare}$&\parbox{1cm}{$|F(\pi)|^2$\vglue-3mm}\\
\km GeV$^2$\km& nb/GeV$^2$    &   nb            &    &
\km GeV$^2$\km& nb/GeV$^2$    &   nb            &    &
\km GeV$^2$\km& nb/GeV$^2$    &   nb            &    \\
 \hline
\input{./table2a_3.inc}
\hline
\end{tabular}
\caption{Cross sections $\sigma_{\pi\pi\gamma}$,  $\sigma_{\pi\pi}^{bare}$  and pion form factor $|F_\pi|^2$
in bins of 0.01 GeV$^2$. The squared mass values are given at the bin center. 
The $\sigma_{\pi\pi\gamma}$ cross section is given as a function of $M^2_{\pi\pi}$.
The $\sigma_{\pi\pi}^{bare}$ cross section and $|F_\pi^2|$ are given as function of $(M^0_{\pi\pi})^2$, see text. The error given is the statistical uncertainty. For $\sigma_{\pi\pi\gamma}$, the second 
error gives the total systematic uncertainty.}
\label{tab:results}}
\end{table}
\end{landscape}

\section{Comparison with previous KLOE results}
\label{sec:CompKLOE}
\begin{figure}
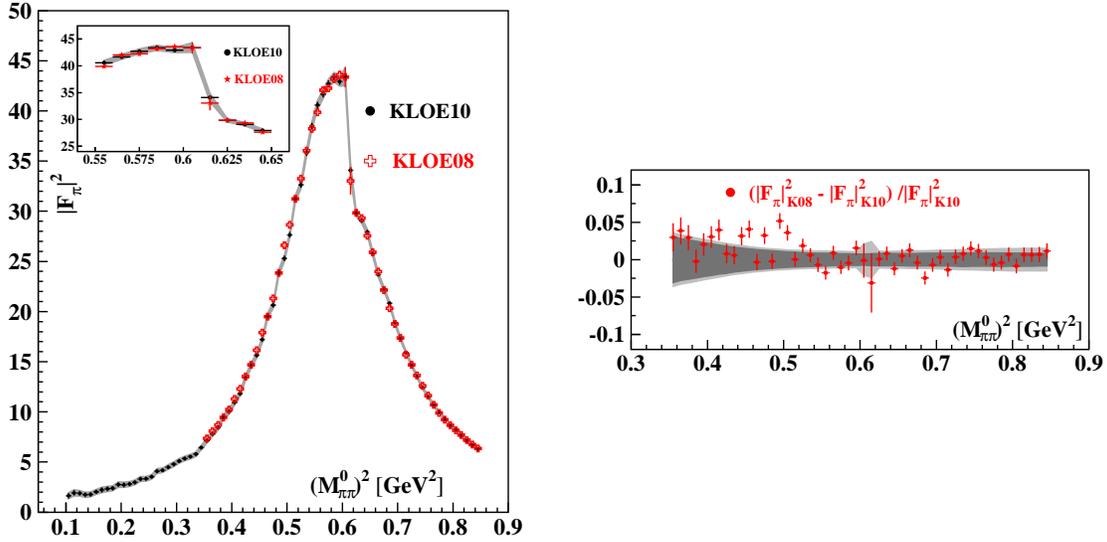

\centerline{\lfigbox fpi_kloe08_09;6.9;\kern.5cm\lfigbox dfpi_kloe08_09;7.2;}
\vglue-0.3cm
\caption{Comparison of the present result, KLOE10, with the previous KLOE result, KLOE08~\cite{plb670}. Left:  Pion form factor $|F_\pi|^2$. Right: Fractional difference between KLOE08 and
 KLOE10 results. The dark (light) gray band gives
  the statistical (total) error for the present result. 
Errors on KLOE08 points contain the combined statistical
  and systematic uncertainty.}
\label{fig:fpi09_08}
\end{figure}
We compare in 
Fig.~\ref{fig:fpi09_08} our
present result for the pion form factor in the range of $0.35 < (M_{\pi\pi}^0)^2 < 0.85$ GeV$^2$ with the result of the previous
KLOE measurement~\cite{plb670}. We stress that data sets have been
obtained at different operating conditions of the DA$\Phi$NE
collider, and different selection cuts in acceptance were used. 
Also the analysis procedures are different since in the previous KLOE analysis the radiated photon was not detected. 
An excellent agreement is
found for $(M_{\pi\pi}^0)^2 > 0.5$ GeV$^2$, while below the new result is lower by few
percent. This is reflected also in the evaluation of the dispersion
integral, see Eq.~\ref{amuint}, between
$0.35$ and $0.85$ GeV$^2$. The new result gives a value of 
$\Delta a_\mu^{\pi\pi}$ which is lower by $(0.8\pm 0.9)$\% (see
Table~\ref{tab:compKLOE08_10}). The experimental systematic precision in the overlapping
 range of $(M_{\pi\pi}^0)^2$ is comparable in both measurements. Systematic effects are 
independent in the two cases except for the uncertainties related to the radiator function, the vacuum polarization and the luminosity measurement, which are identical. 
\begin{table}[htb!]
\begin{center}
 \renewcommand{\arraystretch}{1.5}
  \begin{tabular}{l|c}
     & $\Delta a_\mu^{\pi\pi} (0.35 - 0.85\; {\rm GeV^2}) \times 10^{-10} $ \\
    \hline \hline
    KLOE10 (This work) & $376.6 \pm 0.9_{\rm stat} \pm 2.4_{\rm exp} \pm 2.3_{\rm th}$ \\
    KLOE08~\cite{plb670} & $379.6 \pm 0.4_{\rm stat} \pm 2.4_{\rm exp} \pm 2.2_{\rm th}$ \\
    \hline
  \end{tabular}
\caption{\label{tab:compKLOE08_10}$\Delta a_\mu^{\pi\pi}$ values in the range
  $0.35 - 0.85\; {\rm GeV^2}$.}
\end{center}
\end{table}

Constructing the weighted average of the two measurements we 
evaluate the dispersion integral from $0.1$ to $0.95$  
GeV$^2$, using the method of~\cite{EidJeg95}. Separating out the uncertainties common to both measurements, we obtain
\begin{equation}
\Delta a_\mu^{\pi\pi}(0.1-0.95\; {\rm GeV^2}) = (488.6 \pm
  5.3_{\rm indep.} \pm 2.9_{\rm common}) \times 10^{-10}~.
  \label{eqn:amupipi_pop2}
\end{equation}
The combined fractional total error of $\Delta
a_\mu^{\pi\pi}$ in this range is 1.2\%. 

\section{Comparison with results from the CMD-2, SND and BaBar experiments}
\label{sec:CompOther}
\begin{figure}
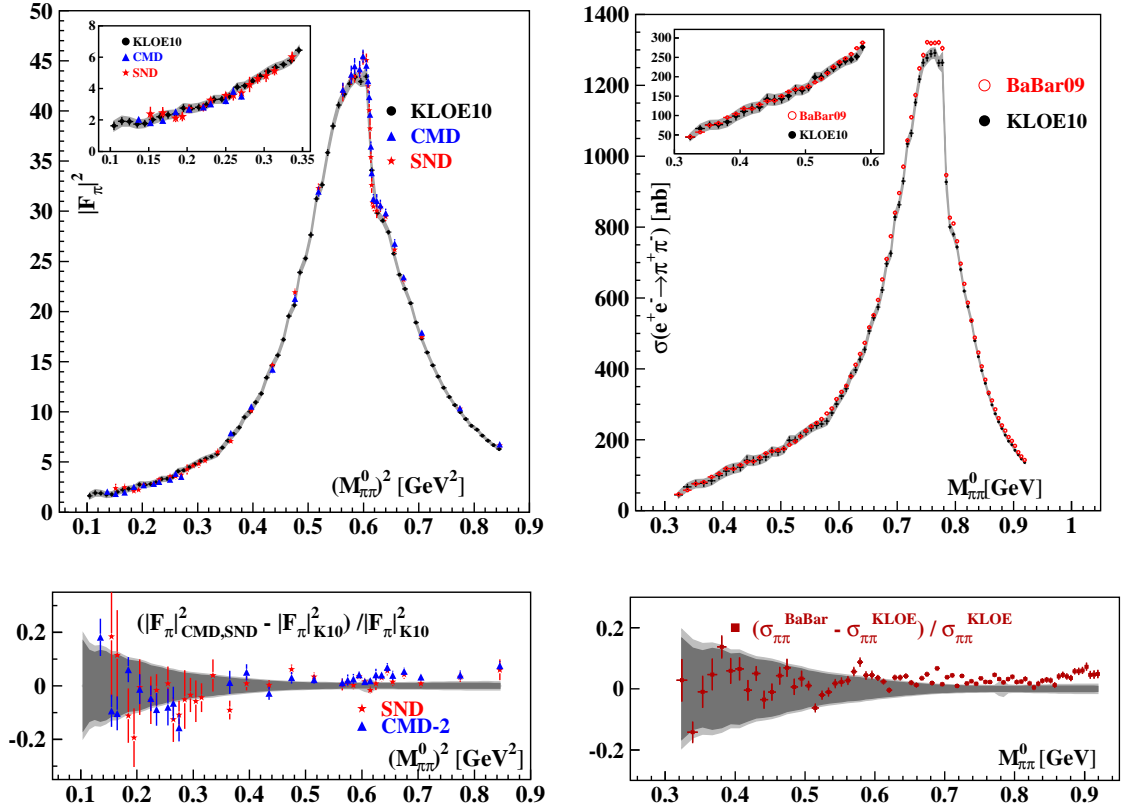

\centerline{\lfigbox fpi_kloe09_snd_cmd;6.9;\kern.5cm\lfigbox
  spp_kloe09_babar09;6.9;}
\centerline{\lfigbox dfpi_kloe09_cmd_snd;7.2;\kern.4cm\lfigbox
  diff_spp_kloe09_babar_int;7.2;}
\vglue-0.3cm
\caption{Top left: $|F_\pi|^2$ from CMD-2~\cite{CMD2, CMD2_2},
  SND~\cite{SND} and the present KLOE result as function of $(M^0_{\pi\pi})^2$. Bottom left: Fractional
  difference between CMD-2 or SND and KLOE. Top right: $\sigma^{\rm bare}_{\pi\pi}$ from
BaBar~\cite{:2009fg} and the new KLOE result as function of $M^0_{\pi\pi}$. Bottom right: Fractional
  difference between BaBar and KLOE. CMD-2, SND and BaBar data points have the total 
uncertainty attached. The
dark (light) band in the lower plots shows statistical (total) error of the KLOE result.}
\label{fig:fpi09_other}
\end{figure}

In Fig.~\ref{fig:fpi09_other}, the new KLOE result is compared with the results from the energy
scan experiments CMD-2~\cite{CMD2, CMD2_2} and SND~\cite{SND} in
Novosibirsk and the result obtained from the BaBar
experiment at SLAC~\cite{:2009fg}, which uses the ISR method. Whenever 
several data points fall in one KLOE bin of $0.01$ GeV$^2$, 
the values are statistically
averaged. Fig.~\ref{fig:fpi09_other}, left, shows the comparison of
$|F_\pi|^2$ obtained by the CMD-2 and SND collaborations with the present
KLOE result. On the $\rho$-peak and above, the agreement with the SND result is rather good, while the result from the CMD-2 collaboration is slightly higher than the new KLOE measurement, confirming the observation already reported in the previous
KLOE publication~\cite{plb670}. 
Below the $\rho$-peak, all three experiments are in agreement within
uncertainties. Fig.~\ref{fig:fpi09_other}, right, shows the present KLOE and the BaBar 
result for the bare cross section as a function of
$M^0_{\pi\pi}$. The fractional difference between BaBar and KLOE
results is shown together with the statistical and total fractional KLOE errors. The two results are in agreement within errors below 0.65 GeV, while above the new BaBar measurement is about 2-3\% higher. 

\section{Conclusions}
We have measured the differential radiative cross section
$ \dd\sigma(e^+e^-\to\pi^+\pi^-\gamma) / \dd M_{\pi\pi}^2 $ in the
interval $0.1 < M^2_{\pi\pi} < 0.85$ GeV$^2$ using 230 pb$^{-1}$ of data
obtained while the DA$\Phi$NE $e^+e^-$ collider was running at $W
\simeq 1$ GeV, 20 MeV below the $\phi$-meson peak. A
systematic uncertainty of $1 \%$ has been reached above $0.4$ GeV$^2$,
rising up to $10 \%$  when approaching $0.1$ GeV$^2$. This increase
is mainly due to the uncertainty in the production mechanism of $\phi$ radiative decays and the uncertainty on the treatment of final state radiation.  

From
this measurement, we have extracted the squared modulus of the pion
form factor in the time-like region, $|F_\pi|^2$, and the bare cross section
for the process $e^+e^- \to \pi^+\pi^-$, $\sigma_{\pi\pi}^{\rm bare}$, in 
  intervals of $0.01$ GeV$^2$ of $(M^0_{\pi\pi})^2$, the squared mass
  of the virtual photon produced in the $e^+e^-$-collision after the
  radiation of a hard photon in the initial state. Our new measurement
  is in good agreement with previous
  KLOE measurements, and reaches down to the dipion production
  threshold. A reasonable agreement has also been found with the results from
  the Novosibirsk experiments CMD-2 and SND, especially at low values
  of $(M^0_{\pi\pi})^2$. Comparing our result with the new result from
  the BaBar collaboration, we have found agreement within errors below     
 $0.4$ GeV$^2$, while above the BaBar result is higher by 2-3\%. 

Evaluating
  the dispersion integral for the dipion contribution to the
muon magnetic moment anomaly, $\Delta a_\mu^{\pi\pi}$, in the range between
$0.1$ and $0.85$ 
GeV$^2$ we have found
\begin{displaymath}
\Delta a_\mu^{\pi\pi}(0.1-0.85\; {\rm GeV^2}) = (478.5 \pm
  2.0_{\rm stat} \pm 5.0_{\rm exp} \pm 4.5_{\rm th}) \times 10^{-10},
\end{displaymath}   
confirming the discrepancy between the SM evaluation for
$a_\mu$ and the experimental value measured by the Muon g-2 collaboration
at BNL.

Combining our result with the previous KLOE results, we have calculated $\Delta a_\mu^{\pi\pi}$ in the range $0.1 < M^2_{\pi\pi} < 0.95$  GeV$^2$ obtaining
 \begin{displaymath}
\Delta a_\mu^{\pi\pi}(0.1-0.95\; {\rm GeV^2}) = (488.6 \pm
  6.0) \times 10^{-10}.
\end{displaymath}   
The KLOE experiment covers $\sim 70\%$ of the leading order hadronic contribution to the muon anomaly with $\sim 1\%$ total error. 

\section*{Acknowledgements}
We wish to acknowledge the work of B. Valeriani in the development of the \mbox{$\pi$-e} PID estimator. 
We would like to thank Carlo Michel Carloni Calame, Henryk Czy{\.z},
Fred Jegerlehner, Johann K\"uhn, Guido Montagna, Germ{\'a}n Rodrigo, Olga Shekhovtsova, Thomas Teubner and Sergiy Ivashyn for numerous useful discussions.

We thank the DA$\Phi$NE team for their efforts in maintaining low background 
running conditions and their collaboration during all data-taking. 
We want to thank our technical staff: 
G.F. Fortugno and F. Sborzacchi for their dedicated work to ensure an
efficient operation of 
the KLOE computing facilities; 
M. Anelli for his continuous support to the gas system and the safety of
the
detector; 
A. Balla, M. Gatta, G. Corradi and G. Papalino for the maintenance of the
electronics;
M. Santoni, G. Paoluzzi and R. Rosellini for the general support to the
detector; 
C. Piscitelli for his help during major maintenance periods.
This work was supported in part
by EURODAPHNE, contract FMRX-CT98-0169; 
by the German Federal Ministry of Education and Research (BMBF) contract 06-KA-957; 
by the German Research Foundation (DFG), 'Emmy Noether Programme',
contracts DE839/1-4;
and by the EU Integrated
Infrastructure
Initiative HadronPhysics Project under contract number
RII3-CT-2004-506078.
%

\vfill
\end{document}

%% file: table2a_3.inc
\km0.105\kak0.34\plm0.06\plm0.03\kak44\plm7\kak1.63\plm0.27\km&%
\km0.355\kak2.91\plm0.09\plm0.03\kak301\plm9\kak7.13\plm0.22\km&%
\km0.605\kak18.57\plm0.12\plm0.35\kak1264\plm10\kak43.42\plm0.33\km\\
\km0.115\kak0.49\plm0.06\plm0.03\kak67\plm9\kak1.92\plm0.26\km&%
\km0.365\kak3.12\plm0.09\plm0.04\kak323\plm9\kak7.79\plm0.22\km&%
\km0.615\kak14.95\plm0.11\plm0.34\kak927\plm7\kak34.09\plm0.27\km\\
\km0.125\kak0.53\plm0.07\plm0.03\kak76\plm9\kak1.89\plm0.24\km&%
\km0.375\kak3.38\plm0.09\plm0.03\kak344\plm9\kak8.43\plm0.22\km&%
\km0.625\kak13.59\plm0.10\plm0.08\kak801\plm7\kak29.81\plm0.25\km\\
\km0.135\kak0.54\plm0.07\plm0.03\kak77\plm10\kak1.74\plm0.23\km&%
\km0.385\kak3.78\plm0.09\plm0.04\kak381\plm9\kak9.47\plm0.23\km&%
\km0.635\kak13.70\plm0.10\plm0.08\kak779\plm6\kak29.08\plm0.24\km\\
\km0.145\kak0.59\plm0.08\plm0.04\kak84\plm11\kak1.78\plm0.23\km&%
\km0.395\kak4.06\plm0.09\plm0.04\kak397\plm9\kak10.02\plm0.23\km&%
\km0.645\kak13.38\plm0.10\plm0.08\kak743\plm6\kak27.91\plm0.23\km\\
\km0.155\kak0.67\plm0.08\plm0.04\kak99\plm11\kak2.02\plm0.23\km&%
\km0.405\kak4.32\plm0.09\plm0.04\kak426\plm9\kak10.94\plm0.23\km&%
\km0.655\kak12.79\plm0.10\plm0.07\kak680\plm6\kak25.77\plm0.21\km\\
\km0.165\kak0.78\plm0.09\plm0.03\kak111\plm13\kak2.21\plm0.26\km&%
\km0.415\kak4.70\plm0.09\plm0.04\kak454\plm9\kak11.83\plm0.23\km&%
\km0.665\kak12.13\plm0.09\plm0.07\kak619\plm5\kak23.68\plm0.20\km\\
\km0.175\kak0.83\plm0.09\plm0.03\kak119\plm12\kak2.32\plm0.24\km&%
\km0.425\kak5.29\plm0.09\plm0.04\kak507\plm9\kak13.40\plm0.24\km&%
\km0.675\kak11.79\plm0.09\plm0.07\kak576\plm5\kak22.25\plm0.19\km\\
\km0.185\kak0.88\plm0.08\plm0.03\kak122\plm12\kak2.38\plm0.23\km&%
\km0.435\kak5.82\plm0.09\plm0.05\kak545\plm9\kak14.62\plm0.24\km&%
\km0.685\kak11.47\plm0.09\plm0.07\kak534\plm5\kak20.84\plm0.18\km\\
\km0.195\kak1.01\plm0.09\plm0.03\kak142\plm13\kak2.75\plm0.26\km&%
\km0.445\kak6.17\plm0.09\plm0.04\kak574\plm9\kak15.64\plm0.24\km&%
\km0.695\kak10.91\plm0.09\plm0.07\kak479\plm4\kak18.91\plm0.16\km\\
\km0.205\kak1.04\plm0.09\plm0.03\kak140\plm13\kak2.72\plm0.24\km&%
\km0.455\kak6.83\plm0.09\plm0.05\kak622\plm9\kak17.21\plm0.25\km&%
\km0.705\kak10.45\plm0.08\plm0.06\kak434\plm4\kak17.32\plm0.15\km\\
\km0.215\kak1.07\plm0.09\plm0.03\kak144\plm12\kak2.81\plm0.23\km&%
\km0.465\kak7.61\plm0.10\plm0.05\kak697\plm9\kak19.55\plm0.26\km&%
\km0.715\kak9.98\plm0.08\plm0.06\kak394.9\plm3.4\kak15.92\plm0.14\km\\
\km0.225\kak1.14\plm0.09\plm0.03\kak151\plm11\kak2.97\plm0.22\km&%
\km0.475\kak8.19\plm0.10\plm0.05\kak725\plm9\kak20.64\plm0.26\km&%
\km0.725\kak9.58\plm0.08\plm0.06\kak359.4\plm3.2\kak14.64\plm0.13\km\\
\km0.235\kak1.29\plm0.09\plm0.03\kak167\plm12\kak3.31\plm0.23\km&%
\km0.485\kak9.37\plm0.10\plm0.06\kak828\plm10\kak23.90\plm0.28\km&%
\km0.735\kak9.30\plm0.08\plm0.06\kak328.5\plm3.0\kak13.53\plm0.12\km\\
\km0.245\kak1.32\plm0.09\plm0.03\kak165\plm11\kak3.32\plm0.22\km&%
\km0.495\kak9.86\plm0.10\plm0.06\kak863\plm10\kak25.30\plm0.28\km&%
\km0.745\kak8.96\plm0.08\plm0.06\kak298.2\plm2.7\kak12.42\plm0.11\km\\
\km0.255\kak1.41\plm0.08\plm0.03\kak173\plm10\kak3.52\plm0.21\km&%
\km0.505\kak10.84\plm0.11\plm0.07\kak930\plm10\kak27.65\plm0.29\km&%
\km0.755\kak8.71\plm0.07\plm0.05\kak272.9\plm2.4\kak11.49\plm0.10\km\\
\km0.265\kak1.64\plm0.09\plm0.03\kak198\plm11\kak4.10\plm0.22\km&%
\km0.515\kak12.25\plm0.11\plm0.08\kak1035\plm10\kak31.24\plm0.31\km&%
\km0.765\kak8.55\plm0.07\plm0.05\kak250.6\plm2.2\kak10.67\plm0.09\km\\
\km0.275\kak1.67\plm0.08\plm0.03\kak199\plm10\kak4.18\plm0.21\km&%
\km0.525\kak12.79\plm0.11\plm0.08\kak1065\plm10\kak32.64\plm0.31\km&%
\km0.775\kak8.42\plm0.07\plm0.05\kak231.8\plm2.1\kak9.97\plm0.09\km\\
\km0.285\kak1.79\plm0.08\plm0.03\kak211\plm10\kak4.49\plm0.21\km&%
\km0.535\kak14.08\plm0.12\plm0.09\kak1151\plm10\kak35.84\plm0.33\km&%
\km0.785\kak8.29\plm0.07\plm0.05\kak213.2\plm1.9\kak9.27\plm0.08\km\\
\km0.295\kak1.92\plm0.08\plm0.03\kak222\plm10\kak4.78\plm0.21\km&%
\km0.545\kak15.20\plm0.12\plm0.09\kak1217\plm11\kak38.49\plm0.34\km&%
\km0.795\kak8.19\plm0.07\plm0.05\kak196.1\plm1.8\kak8.62\plm0.08\km\\
\km0.305\kak2.02\plm0.09\plm0.03\kak233\plm10\kak5.10\plm0.21\km&%
\km0.555\kak16.06\plm0.12\plm0.09\kak1264\plm11\kak40.59\plm0.34\km&%
\km0.805\kak8.32\plm0.07\plm0.05\kak185.2\plm1.6\kak8.23\plm0.07\km\\
\km0.315\kak2.17\plm0.09\plm0.03\kak241\plm9\kak5.36\plm0.21\km&%
\km0.565\kak16.62\plm0.12\plm0.10\kak1278\plm10\kak41.68\plm0.34\km&%
\km0.815\kak8.29\plm0.07\plm0.05\kak170.2\plm1.5\kak7.64\plm0.07\km\\
\km0.325\kak2.26\plm0.09\plm0.03\kak244\plm9\kak5.53\plm0.21\km&%
\km0.575\kak17.38\plm0.12\plm0.10\kak1289\plm10\kak42.71\plm0.34\km&%
\km0.825\kak8.28\plm0.07\plm0.05\kak157.4\plm1.4\kak7.13\plm0.06\km\\
\km0.335\kak2.38\plm0.09\plm0.03\kak252\plm9\kak5.79\plm0.21\km&%
\km0.585\kak17.85\plm0.12\plm0.10\kak1291\plm10\kak43.38\plm0.34\km&%
\km0.835\kak8.34\plm0.07\plm0.05\kak146.1\plm1.2\kak6.69\plm0.06\km\\
\km0.345\kak2.63\plm0.09\plm0.04\kak276\plm9\kak6.44\plm0.21\km&%
\km0.595\kak18.13\plm0.12\plm0.10\kak1263\plm10\kak42.94\plm0.33\km&%
\km0.845\kak8.45\plm0.07\plm0.05\kak135.9\plm1.1\kak6.28\plm0.05\km\\